\documentclass[twocolumn,english,aps,superscriptaddress]{revtex4-2}
\usepackage[T1]{fontenc}
\usepackage[latin9]{inputenc}
\usepackage{amsmath}
\usepackage{graphicx}

\makeatletter
\usepackage{amsthm}
\usepackage{amsfonts}
\usepackage{bbm}
\usepackage{epsfig}
\usepackage{babel}
\usepackage{color}
\usepackage{framed}
\usepackage{changes}
\usepackage{float}
\usepackage{array}

\usepackage{xcolor}      % 支持颜色
\usepackage{titlesec}    % 自定义标题格式
\usepackage{lineno}
\usepackage{setspace}
\modulolinenumbers[1]

\titleformat{\section}[block]   % block 表示左对齐普通段落式
  {\color{black}\normalfont\bfseries\normalsize} % 红色 + 粗体 + 稍大字号
  {}        % 无编号
  {0pt}     % 标题与文本之间的间距
  {}        % 标题前无额外命令

\titlespacing*{\section}
  {0pt}     % 左边距
  {2ex plus 1ex}  % 上方间距
  {0ex}            % 下方间距

\titleformat{\subsection}[block]
  {\color{black}\normalfont\bfseries\small} % 深灰色、粗体、小字号
  {}
  {0pt}
  {}
\titlespacing*{\subsection}
  {0pt}     % 左边距
  {1ex}  % 上方间距
  {0ex}            % 下方间距

\usepackage{hyperref}
\hypersetup{colorlinks=true,citecolor=red,linkcolor=blue}

\makeatother

\usepackage{babel}
\begin{document}

%\begin{linenumbers}
\title{Black Holes Trapped by Ghosts}
\author{Cheng-Yong Zhang}
\email{zhangcy@email.jnu.edu.cn}

\affiliation{\textit{Department of Physics and Siyuan Laboratory, Jinan University,
Guangzhou 510632, China }}
\author{Yunqi Liu}
\email{yunqiliu@yzu.edu.cn}

\affiliation{\textit{Center for Gravitation and Cosmology, College of Physical
Science and Technology, Yangzhou University, Yangzhou 225009, China}}
\author{Bin Wang}
\email{wang\_b@sjtu.edu.cn}

\affiliation{\textit{Center for Gravitation and Cosmology, College of Physical
Science and Technology, Yangzhou University, Yangzhou 225009, China}}
\affiliation{\textit{School of Aeronautics and Astronautics, Shanghai Jiao Tong
University, Shanghai 200240, China}}
\begin{abstract}
Violent cosmic events, from black hole mergers to stellar collapses, often leave behind highly excited black hole remnants that inevitably relax to equilibrium. The prevailing view, developed over decades, holds that this relaxation is rapidly filtered into a linear regime, establishing linear perturbation theory as the bedrock of black hole spectroscopy and a key pillar of gravitational-wave physics. Here we unveil a distinct nonlinear regime that transcends the traditional paradigm: before the familiar linear ringdown, an intrinsically nonlinear, long-lived bottleneck can dominate the evolution. This stage is controlled by a saddle-node ghost in phase space, which traps the remnant and delays the onset of linearity by a timescale obeying a universal power-law. The ghost imprints a distinctive quiescence-burst signature on the emitted radiation: a prolonged silence followed by a violent burst and a delayed ringdown. Rooted in the bifurcation topology, it extends naturally to neutron and boson stars, echoing a topological universality shared with diverse nonlinear systems in nature. Our results expose a missing nonlinear chapter in gravitational dynamics and identify ghost-induced quiescence-burst patterns as clear targets for future observations.
\end{abstract}
\maketitle

%\section{Introduction}
How does a violently disturbed black hole settle down? 
This deceptively simple question sits at the heart of strong-field gravity. 
Catastrophic cosmic events---black hole mergers, core-collapse supernovae, and dramatic accretion episodes---leave behind black hole remnants so strongly distorted that one might expect their relaxation to be governed by nonlinear chaos. 
Yet for decades, observations and simulations have told a remarkably simple story: like struck bells, they ring down \citep{LIGOScientific:2016aoc,Pretorius:2005gq,Berti:2009kk,Konoplya:2011qq}. 
The initial storm of nonlinear dynamics is rapidly filtered by the black hole horizon, driving the system swiftly into a linear regime. The subsequent relaxation is then well captured by linear perturbation theory and proceeds through a ringdown governed by quasinormal modes (QNMs), followed by a late-time tail. This remarkable simplification is the bedrock of black hole spectroscopy \citep{Berti:2025hly}, a central tool of gravitational-wave astronomy that allows us to probe a black hole from its QNM spectrum \citep{LIGOScientific:2016lio,Isi:2019aib,LIGOScientific:2021sio}, much as atomic spectra reveal the atomic structure. While recent studies have identified higher-order nonlinear corrections to the ringdown \citep{Cheung:2022rbm,Mitman:2022qdl,Khera:2023oyf,Khera:2024bjs}, they remain subdominant. These empirical successes have therefore reinforced the prevailing picture: the story of black hole relaxation is dominated by linear physics.

Yet, this linear simplicity stands in stark contrast to a ubiquitous pattern observed in natural and social systems  \citep{scheffer2009critical,strogatz2018nonlinear,izhikevich2007dynamical,ghil2020physics,scheffer2009early,rietkerk2021evasion}: near a tipping point, dynamics are generically dominated not by linear effects, but by intrinsic nonlinearity encoded in the underlying bifurcation structure in phase space. 
It manifests broadly in systems from Venus flytrap snap \citep{forterre2005venus} and elastic beam buckling \citep{gomez2017critical} to ecosystem collapse \citep{scheffer2001catastrophic} and abrupt climate change \citep{alley2003abrupt}. 
This raises a profound question: Is the linear dominance in black holes, the most extreme objects in nature, truly universal? Or can nonlinearity step out of the shadows and become the leading driver of black hole evolution over macroscopic times?

Here we show that nonlinearity indeed can, unveiling an entirely new evolutionary path in black hole relaxation, characterized by a fundamentally novel quiescence-burst energy emission profile. We demonstrate that in gravitational systems whose equilibrium solution space exhibits a saddle-node bifurcation, a black hole driven beyond the tipping point---where two branches of equilibria meet and mutually annihilate---does not promptly relax through the expected ringdown. Instead, it becomes trapped for a long time in a slowly evolving bottleneck phase, lingering near a ghost that reflects the remnant influence of the annihilated equilibria \citep{strogatz2018nonlinear,scheffer2009critical,izhikevich2007dynamical}. Using high-precision numerical simulations, supported by asymptotic analysis, we find that the bottleneck lifetime obeys a universal power-law $t_{b}\propto\epsilon^{-1/4}$, where $\epsilon$ measures the deviation from the tipping point. This scaling is an intrinsic nonlinear consequence of the saddle-node ghost and inaccessible to linear analysis. Our results provide the first evidence that intrinsic nonlinearity can dominate the long-time evolution of a black hole, reshaping the classical relaxation picture not as a small correction to linear theory, but as an essentially new regime. 

This new relaxation path predicts that the aftermath of a black hole merger or accretion event will not always be an immediate ringdown. If the remnant is driven beyond the tipping point, it enters a prolonged silent phase where energy emission is suppressed by orders of magnitude. This quiescence is broken by a delayed violent burst of radiation that finally transitions into a standard ringdown signal. This distinctive quiescence-burst pattern stands beyond the reach of linear theory, and challenges the canonical picture of post-merger evolution which expects a prompt transition to ringdown.

The power of this discovery lies in its profound universality. Arising from the bifurcation topology of the phase space \citep{strogatz2018nonlinear}, the bottleneck transcends the microphysics distinguishing different compact objects. Black holes, neutron stars, and boson stars that share this topology will be governed by the same ghost-induced dynamics. This topological universality reveals a common language between strong-field gravity and diverse nonlinear systems in nature \citep{scheffer2009critical,strogatz2018nonlinear,izhikevich2007dynamical,ghil2020physics,scheffer2009early,rietkerk2021evasion,forterre2005venus,gomez2017critical,scheffer2001catastrophic,alley2003abrupt}, elevating the quiescence-burst signature from a new emission pattern to a broadly applicable diagnostic of fundamental physics.

\begin{figure}[h]
\begin{centering}
\includegraphics[width=0.99\linewidth]{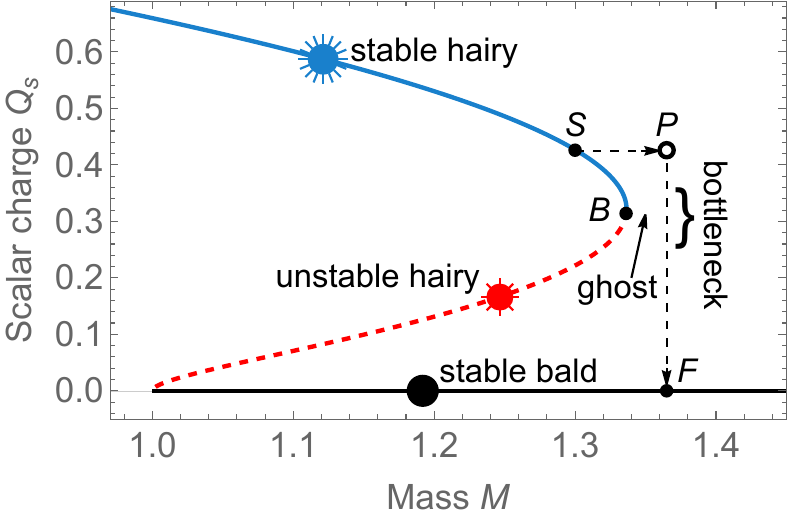}
\par\end{centering}
{\footnotesize\caption{{\footnotesize\protect\label{fig:static}\textbf{Bifurcation structure of
the static black hole solution space.} The scalar charge $Q_s$ is plotted against the total mass $M$, with the electric charge fixed at $Q=1$ to define the unit length scale. Three equilibrium branches exist: the stable hairy branch (solid blue), the unstable hairy branch (dashed red), and the stable bald branch (black line, $Q_s=0$). The two hairy branches annihilate at the tipping point $B$, forming a saddle-node bifurcation. A representative dynamical pathway is indicated by the initial stable hairy state $S$, the nonequilibrium post-perturbation state $P$, and the final bald state $F$. Results are shown for coupling $\lambda=100$, but the bifurcation structure is generic for other $\lambda$.}}
}{\footnotesize\par}
\end{figure}

\section{Saddle-node bifurcation and ghost}
To expose the ghost-induced bottleneck in a clean yet representative form, we consider spherically symmetric black holes in Einstein-Maxwell-scalar theory with Lagrangian density $\mathcal{L}=R-2\nabla_{\mu}\phi\nabla^{\mu}\phi-f(\phi)F_{\mu\nu}F^{\mu\nu}$, where $\phi$ is a real scalar field coupled to the Maxwell invariant through $f=e^{\lambda\phi^{4}}$. 
To handle equilibrium solutions and their dynamical evolution within a unified framework, we employ the Painlev\'e-Gullstrand coordinates, which remain regular across the horizon (see Methods). 
Fig. \ref{fig:static} shows the set of equilibria.
Alongside the stable bald branch, two distinct branches of hairy solutions emerge: one linearly stable and one unstable.
These branches meet and annihilate at the tipping point $B$ (mass $M_*$), forming the classic saddle-node bifurcation familiar from nonlinear dynamics \citep{strogatz2018nonlinear}. 
Crucially, for $M > M_*$, the hairy equilibria vanish, yet their influence persists as a remnant "ghost" in phase space that traps  nearby evolutionary paths. To probe this structure dynamically, we inject mass into a stable hairy black hole $S$ via an ingoing matter pulse of strength $p$. A threshold $p_*$ separates two dynamical regimes. Subcritical perturbations ($p<p_*$) merely trigger a standard ringdown to a nearby hairy state. In sharp contrast, supercritical perturbations drive the system to a nonequilibrium state $P$ with $M>M_*$ which inevitably evolves to a bald state $F$,  but the path can be trapped by the ghost.

\begin{figure}[h]
\begin{centering}
\includegraphics[width=0.99\linewidth]{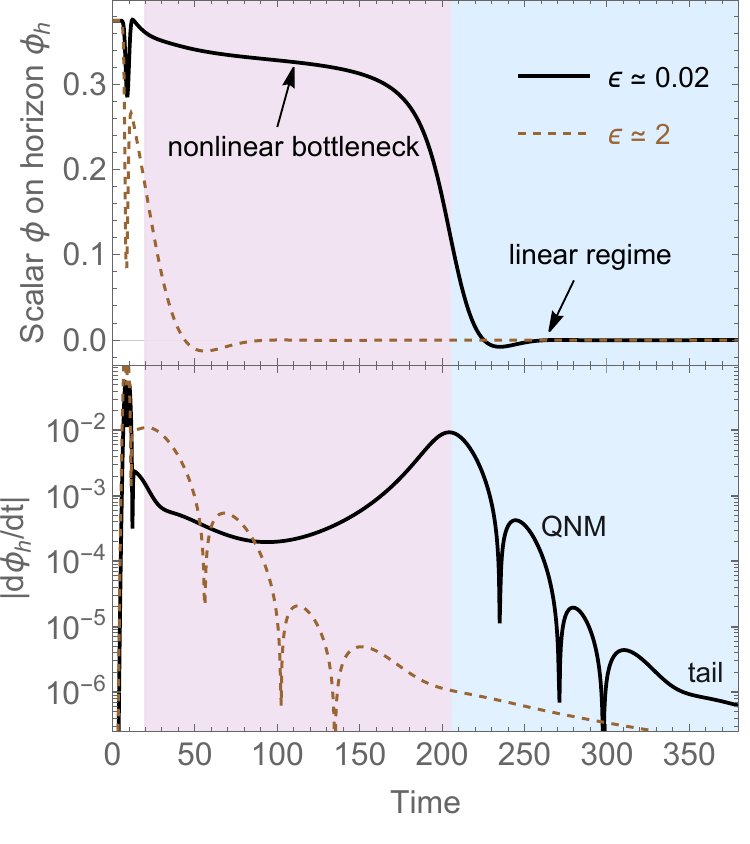}
\par\end{centering}
{\footnotesize\caption{{\footnotesize\protect\label{fig:bottleEvo} 
\textbf{Dynamical signature of the nonlinear bottleneck.} Top: Time evolution of the scalar field at the horizon, $\phi_h(t)$. Bottom: The time derivative $|d\phi_h/dt|$, highlighting the decay rates. All simulations start from the stable hairy solution $S$ with $M=1.3$ (see Fig. \ref{fig:static}), perturbed by an ingoing Gaussian scalar pulse $\delta\phi = -p e^{-(r-16)^2/4}$. The threshold is $p_* \simeq 0.012753$. The far-supercritical case ($\epsilon \simeq 2$, dashed brown) transitions promptly to linear ringdown. The near-critical evolution ($\epsilon\simeq 0.02$, black solid) exhibits a prolonged bottleneck (purple) prior to ringdown (blue). The bottleneck time $t_b$ is identified by  the peak in $|d\phi_h/dt|$ after the initial burst (interface between purple and blue regions).}}
}{\footnotesize\par}
\end{figure}

\section{Bottleneck dynamics and power-law scaling}
Fig. \ref{fig:bottleEvo}  illustrates this trap,  contrasting evolutions parameterized by the relative deviation $\epsilon=(p-p_{*})/p_{*}$.
For a strong perturbation far above threshold ($\epsilon\simeq2$), the black hole quickly absorbs the scalar field, and settles into the bald state with dynamics well described by linear QNMs. 
However, a near-critical perturbation
($\epsilon\simeq0.02$) triggers a qualitatively new behavior. 
Instead of a prompt decay, the system is held near the tipping point by the ghost, remaining in a slowly evolving bottleneck for hundreds of dimensionless time units.
The bottom panel  highlights the nonlinear nature of this bottleneck: while a linear ringdown follows a straight-line envelope on a logarithmic plot \citep{Berti:2009kk,Konoplya:2011qq,Berti:2025hly}, the bottleneck curve strongly deviates from this linearity (shaded purple). Only after escaping this nonlinear bottleneck does the system enter the familiar linear regime (shaded blue).

\begin{figure}[h]
\begin{centering}
\includegraphics[width=0.99\linewidth]{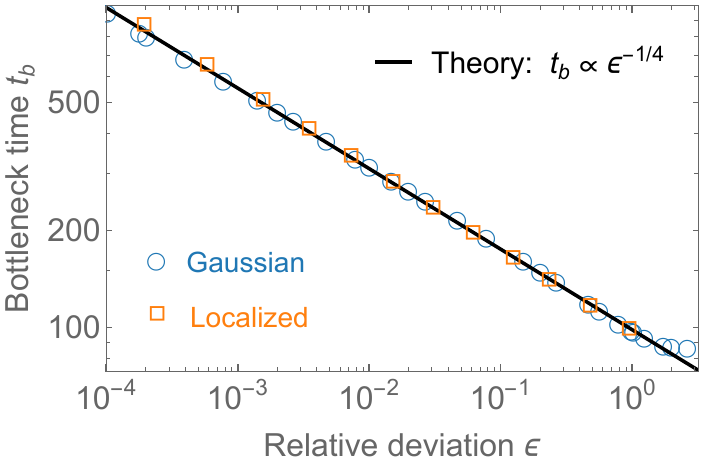}
\par\end{centering}
{\footnotesize\caption{{\footnotesize\protect\label{fig:tbscaling} 
\textbf{Universal scaling of
the bottleneck time.} The bottleneck time $t_{b}$ follows a universal
power-law $t_{b}\propto\epsilon^{-1/4}$ (black line), where 
$\epsilon$ measures the relative deviation from threshold.
Two families of initial perturbations are shown: Gaussian pulses $\delta\phi=-pe^{-(r-16)^{2}/4}$
(blue circles) and compact localized pulses $\phi=pe^{-\frac{r}{r-r_{1}}-\frac{r}{r_{2}-r}}\left(\frac{r}{r-r_{1}}\right)^{5}\left(\frac{r}{r_{2}-r}\right)^{5}$
with $r_{1}=7$ and $r_{2}=8$ (orange squares). We have simulated the evolutions with other coupling  $\lambda$ and initial hairy black holes, and found this $-1/4$ power-law holds universally, even up to $\epsilon\sim1$, underscoring the robustness and potential
astrophysical relevance.}}
}{\footnotesize\par}
\end{figure}

Fig. \ref{fig:tbscaling} reveals that the bottleneck time $t_b$, defined as the delay before the final transition to ringdown, obeys a universal power-law with the relative deviation,
\begin{equation}
t_{b}\propto\epsilon^{-1/4}.\label{eq:tb}
\end{equation}
This scaling is robust across different initial profiles and fundamentally distinguishes the bottleneck from other critical phenomena in gravity, such as type I critical collapse \citep{Gundlach:2007gc} and black hole scalarization \citep{Zhang:2021nnn}. Those processes follow a logarithmic law  $t \propto \ln\epsilon^{-1}$, the hallmark of a system escaping an unstable equilibrium via a linear eigenmode. The bottleneck, in contrast, lacks such an underlying equilibrium; the system instead navigates a ghost region in phase space where the hairy solution has ceased to exist.  
The power-law (\ref{eq:tb}) is therefore a smoking gun of intrinsically nonlinear dynamics.

\section{Topological origin and university}
To uncover the physical origin of this $-1/4$ power-law, we must transcend standard linear perturbation theory,  which faces a structural impasse here. In the supercritical regime, the vanishing of the hairy equilibrium leaves no suitable background for linear expansion. Linearizing at the bifurcation point is equally futile: the spectrum is dominated by a zero mode with vanishing eigenfrequency. The standard ansatz \(e^{-i\omega t}\) renders this mode strictly static \citep{Berti:2009kk,Konoplya:2011qq,Berti:2025hly}, making linear theory blind to the slow drift that characterizes the bottleneck.

To isolate the slow dynamics, we turn to employ the multiscale method to perform a center manifold reduction \citep{Keener2000,kuznetsov1998elements}, and develop a systematic asymptotic analysis near the bifurcation solution $\phi_*(r)$ at the tipping point $B$ in Fig. \ref{fig:static}  (see Methods). Central to this analysis is the corresponding zero mode $\phi_0(r)$, whose amplitude is promoted to a dynamical variable $A(T)$ evolving on a slow time $T=\varepsilon^{\eta}t$.
We expand the scalar field around the bifurcation solution as $\phi\to\phi_{*}+\delta \phi$, where
\begin{equation}
\delta\phi(t,r)=\varepsilon^{\gamma}A(T)\phi_{0}(r)+\text{higher order terms},\label{eq:phi0ln}
\end{equation}
with analogous expansions for the metric functions. Here $\varepsilon=(M-M_{*})/M_{*}$ serves as the control parameter measuring the mass deviation from the tipping point. It is induced by the perturbation and is proportional to $\epsilon$. A nontrivial slow evolution requires a balance between the inertial term  $\partial_{t}^{2}\delta\phi$, the leading nonlinearity $\delta\phi^{2}$, and the external driving force set by  $\varepsilon$.  This dominant balance uniquely fixes $\gamma=1/2$ and $\eta=1/4$. 
These exponents arise in snap-through instabilities of elastic beams \citep{gomez2017critical,radisson2023elastic}, indicating a shared dynamical structure that underlies these vastly different physical settings.

By enforcing the Fredholm solvability condition \citep{Keener2000}, the complex field dynamics is projected onto a remarkably simple effective equation for the physical amplitude $\varLambda=\varepsilon^{1/2}A$ of the zero mode:
\begin{equation}
d^{2}\varLambda/dt^{2}=-\mu\varepsilon-\beta\varLambda^{2},\label{eq:Lt0}
\end{equation}
where $\mu$ and $\beta$ are positive constants determined by the bifurcation solution. This equation reveals the mechanical basis of the bottleneck: the system behaves as a particle moving in a weakly tilted cubic potential $V=\mu\varepsilon\varLambda+\beta\varLambda^{3}/3$.
The equilibria correspond to the extrema of $V$, reproducing the saddle-node bifurcation where stable and unstable branches annihilate at $\varepsilon=0$. 
While the equilibrium vanishes for $\varepsilon>0$, the potential landscape remains nearly flat in its vicinity. This flatness preserves a dynamical memory of the annihilated state, known in dynamical systems as a ghost \citep{strogatz2018nonlinear,scheffer2009critical,izhikevich2007dynamical}, which traps the particle in a prolonged, slowly evolving bottleneck phase  before it finally accelerates away.
The universal scaling is an immediate consequence of this picture. In the slow phase, the zero mode amplitude scales as $\varLambda\sim\varepsilon^{1/2}$. Balancing the inertial term $d^{2}\varLambda/dt^{2}\sim\varLambda/t_{b}^{2}$ with the potential force  $-\partial V/\partial\varLambda\sim\varepsilon$ yields the power-law $t_{b}\propto\varepsilon^{-1/4}$, analytically confirming the $\epsilon^{-1/4}$ scaling found in simulations.

The reduced dynamics demonstrates that the bottleneck is controlled fundamentally by the saddle-node ghost acting on a system with inertia. The essential ingredients are simply the annihilation of equilibrium branches and the lingering dynamical influence of the ghost. Its origin is therefore topological \citep{strogatz2018nonlinear,izhikevich2007dynamical,kuznetsov1998elements}: the mechanism depends neither on the microscopic details of the field equations nor on spatial symmetry. Our analysis identifies the black hole bottleneck precisely with this generic slow passage, a phenomenon that is intrinsically nonlinear and invisible to linear theory.

%\section{Universality in the extreme}

This topological origin implies that the phenomenon should persist far beyond our specific setup. Indeed, such bifurcations are widespread in gravitational physics. They appear in spherical and rotating black holes across a variety of theories \citep{Corelli:2022pio,Emparan:2008eg,Figueras:2017zwa,Janik:2017ykj,Liu:2025eve,Zhang:2023qtn,Doneva:2022yqu,Doneva:2022ewd}. In Einstein-dilaton-Gauss-Bonnet gravity, for instance, the saddle-node sets a minimal black hole mass \citep{Corelli:2022pio}, implying that Hawking evaporation would encounter a bottleneck as the mass approaches this bound.  In higher dimensions, the bifurcation between black holes and black rings \citep{Emparan:2008eg} suggesting  a ghost may affect constraints on cosmic censorship \citep{Figueras:2017zwa}. Even in the holographic framework, this topology maps to spinodal instabilities, governing the non-equilibrium dynamics of phase separation in the dual quantum matter \citep{Janik:2017ykj}.
The same topology also governs matter-filled compact objects. In neutron stars, the  transition to deconfined quark matter generates a bifurcation in the mass-radius relation \citep{Yagi:2016bkt,Fischer:2017lag,Bauswein:2018bma,Doneva:2022ewd}, creating a stage for slow ghost dynamics to encode phase transitions deep within the core.  Boson stars,  prominent
dark matter candidates \citep{Liebling:2012fv,Cardoso:2019rvt,Marks:2025jpt}, also exhibit branches terminating at a tipping point, indicating that ghosts could control their dynamical response to mergers or accretion. 
Regardless of the physical constituents, from vacuum curvature to dense matter, the dynamics are driven by a shared ghost topology which should imprint a characteristic signature on the emitted radiation, be it electromagnetic or gravitational.

\begin{figure}[h]
\begin{centering}
\includegraphics[width=0.99\linewidth]{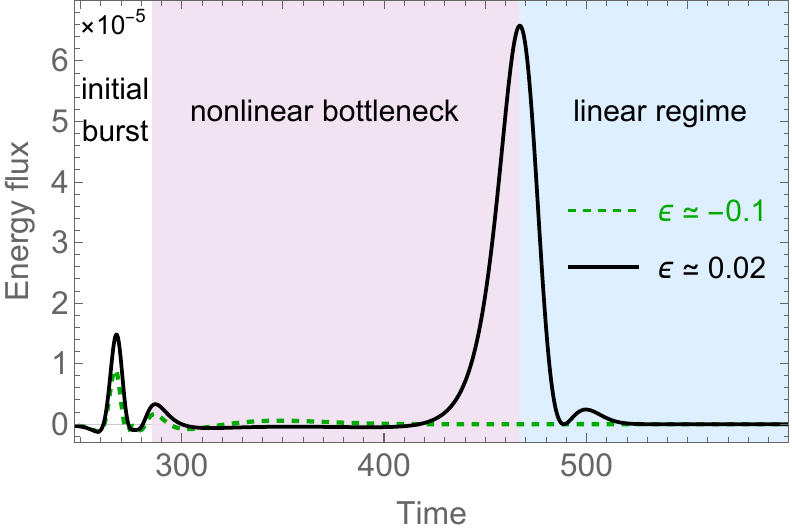}
\par\end{centering}
{\footnotesize\caption{{\footnotesize\protect\label{fig:flux} \textbf{The   quiescence-burst
energy emission signature of the nonlinear bottleneck.} Energy flux
measured at radius $r=200$ as a function of time. The curve for $\epsilon\simeq0.02$ has
three stages: the initial burst, the nonlinear bottleneck phase (shaded
purple) and the linear regime (shaded blue). While a typical perturbation
($\epsilon\simeq-0.1$,  dashed green) radiates energy immediately, and
has only two stages: the initial burst and the linear regime.}}
}{\footnotesize\par}
\end{figure}

\section{Quiescence-burst energy emission signature}

To quantify this distinctive emission profile, we compute the outgoing energy flux $-dm/dt$ across a large-radius sphere, where $m$ is the Misner-Sharp quasi-local mass \citep{Hayward:1994bu}. This flux serves as a proxy for the gravitational-wave emission expected from realistic, asymmetric events. 
Fig. \ref{fig:flux} contrasts two representative runs.  The  dashed green curve ($\epsilon\simeq-0.1$) shows the standard pattern: after an initial burst, the energy is promptly radiated away through linear ringdown. 
In contrast, the energy emission for a near-critical perturbation ($\epsilon\simeq0.02$) exhibits a distinct three-stage sequence: an initial burst, followed by a prolonged phase of dynamical quiescence, where the energy flux is suppressed by orders of magnitude for hundreds of dimensionless time units (shaded purple), and finally a transition to the linear regime (shaded blue). 
The quiescence corresponds precisely to the bottleneck, where the system stores its excess energy nonlinearly, instead of radiating it away. The stored energy is only released catastrophically once the bottleneck is crossed, producing a characteristic quiescence-burst signature that is much more luminous than the usual ringdown emission.

As shown in Fig. \ref{fig:tbscaling}, the power-law persists even for  large deviations  ($\epsilon\sim1$), indicating that the bottleneck is not a fine-tuned effect, but a robust astrophysical phenomenon. Its robustness is rooted in the topological nature of the ghost \citep{strogatz2018nonlinear,kuznetsov1998elements}, which organizes slow dynamical flows even in noisy environments \citep{koch2024ghost}. 
This mechanism predicts a measurable silent delay between an initial dynamical trigger, such as a merger or accretion event, and the ringdown signal. 
For stellar-mass black holes, the scaling yields a delay of $\sim 10^{-1}$ seconds; for a supermassive black hole ($10^5M_\odot$), the silence spans roughly $10^2$ seconds. 
Such delayed-onset ringdowns can be detected by ground-based (LIGO-Virgo-KAGRA, ET, CE) \citep{KAGRA:2013rdx,ET:2019dnz,Reitze:2019iox}  and space-based  (LISA, Taiji, TianQin) \citep{LISA:2024hlh,Ruan:2018tsw,TianQin:2015yph} detectors in their respective regimes.  
Beyond verifying the bottleneck itself, such observations would serve as a sharp diagnostic for the topology of the solution space, constraining exotic compact objects and alternative theories that host similar bifurcations.

By bridging strong-field gravity with the ubiquitous nonlinear behaviors observed in diverse natural systems \citep{scheffer2009critical,strogatz2018nonlinear,izhikevich2007dynamical,ghil2020physics,scheffer2009early,rietkerk2021evasion,forterre2005venus,gomez2017critical,scheffer2001catastrophic,alley2003abrupt}, this work establishes the ghost-induced bottleneck as a new paradigm for black hole dynamics.  
This discovery transcends the traditional linear picture of black hole relaxation, revealing that nonlinearity can step out of the shadow of linear effects to become the dominant driver of evolution for macroscopic times. It offers a powerful new lens for probing fundamental physics in the dynamics of gravitational systems, demonstrating that the journey to equilibrium is not always direct---it can be silent, then violent.

\section{Acknowledgments}

We thank Peng-Cheng Li, Zhen Zhong, Yu-Peng Zhang, Lin Chen for helpful
discussions. 
This work is supported by National Key R$\&$D Program of China No. 2023YFC2206703, and Natural Science Foundation
of China (NSFC) under Grant No. 12375048, 12375056.

\bibliographystyle{apsrev4-2ideal}
\bibliography{bottleneck}

\section{Methods}

\subsection{Numerical setup for dynamical simulation}

The Einstein equations in the specific Einstein-Maxwell-scalar theory
we considered are
\begin{equation}
R_{\mu\nu}-\frac{1}{2}Rg_{\mu\nu}=2\left(T_{\mu\nu}^{\phi}+f(\phi)T_{\mu\nu}^{A}\right),
\end{equation}
where the stress-energy tensor for the scalar field $\phi$ and Maxwell
field $A_{\mu}$ are respectively:
\begin{align}
T_{\mu\nu}^{\phi}= & \partial_{\mu}\phi\partial_{\nu}\phi-\frac{1}{2}g_{\mu\nu}\nabla_{\rho}\phi\nabla^{\rho}\phi,\\
T_{\mu\nu}^{A}= & F_{\mu\rho}F_{\nu}^{\ \rho}-\frac{1}{4}g_{\mu\nu}F_{\rho\sigma}F^{\rho\sigma},\nonumber 
\end{align}
 with field strength $F_{\mu\nu}=\partial_{\mu}A_{\nu}-\partial_{\nu}A_{\mu}$.
The Maxwell field satisfies
\begin{equation}
\nabla_{\mu}\left(f(\phi)F^{\mu\nu}\right)=0.\label{eq:eqMaxwell}
\end{equation}
 The scalar field satisfies 
\begin{equation}
\nabla_{\mu}\nabla^{\mu}\phi=\frac{\dot{f}}{4}F_{\mu\nu}F^{\mu\nu},\label{eq:eqScalar}
\end{equation}
where $\dot{f}=\frac{df(\phi)}{d\phi}$. 

To study the full nonlinear evolution of a black hole under perturbation,
we take the Painlev\'e-Gullstrand (PG) coordinates, which remain regular
across the horizon. The line element in PG coordinates reads
\begin{equation}
ds^{2}=-\left(1-\zeta^{2}\right)\alpha^{2}dt^{2}+2\zeta\alpha dtdr+dr^{2}+r^{2}d\Omega^{2}.\label{eq:PG}
\end{equation}
where $d\Omega^{2}=d\theta^{2}+\sin^{2}\theta d\varphi^{2}$, and
$\zeta,\alpha$ are metric functions depending on $(t,r)$. The normal
vector to the time slice $n^{\mu}=(\alpha^{-1},-\zeta,0,0)$.

Taking the gauge $A_{\mu}=(A,0,0,0)$, the Maxwell equations (\ref{eq:eqMaxwell})
give $\partial_{r}A=\frac{Q\alpha}{r^{2}f}$ where $Q$ is a constant
representing the black hole electric charge. We set $Q=1$ as our
unit length scale throughout our analysis. Introducing an auxiliary
variable $\Pi=n^{\mu}\nabla_{\mu}\phi=\frac{1}{\alpha}\partial_{t}\phi-\zeta\partial_{r}\phi$, the Einstein
equations turn to 
\begin{align}
\partial_{r}\zeta= & \frac{r}{2\zeta}\left((\partial_{r}\phi)^{2}+\Pi^{2}\right)+\frac{Q^{2}}{2r^{3}\zeta f}-\frac{\zeta}{2r}+r\Pi\partial_{r}\phi,\label{eq:zetadr}\\
\partial_{r}\alpha= & -\frac{r\alpha\Pi\partial_{r}\phi}{\zeta},\label{eq:alphadr}\\
\partial_{t}\zeta= & \frac{r\alpha}{\zeta}\left(\Pi+\zeta\partial_{r}\phi\right)\left(\Pi\zeta+\partial_{r}\phi\right),\label{eq:zetadt}
\end{align}
and the scalar field equation gives 
\begin{align}
\partial_{t}\phi & =\alpha\left(\Pi+\zeta\partial_{r}\phi\right),\label{eq:phit}\\
\partial_{t}\Pi & =\frac{\partial_{r}\left[\left(\Pi\zeta+\partial_{r}\phi\right)\alpha r^{2}\right]}{r^{2}}+\frac{\alpha}{2}\frac{Q^{2}\dot{f}}{r^{4}f^{2}}.\label{eq:Pt}
\end{align}

\textbf{Static solution}. 
A static solution of this system satisfies 
\begin{align}
0= & \partial_{r}\alpha_{s}-r\alpha_{s}(\partial_{r}\phi_{s})^{2},\label{eq:alphas}\\
0= & \partial_{r}\zeta_{s}+\frac{\zeta_{s}}{2r}-r\frac{1-\zeta_{s}^{2}}{2\zeta_{s}}\left(\partial_{r}\phi_{s}\right)^{2}-\frac{Q^{2}}{2r^{3}\zeta_{s}f_{s}},\label{eq:zetas}\\
0= & \partial_{r}^{2}\phi_{s}+\frac{\left(\zeta_{s}^{2}-2+\frac{Q^{2}}{r^{2}f_{s}}\right)\frac{\partial_{r}\phi_{s}}{r}-\frac{Q^{2}\dot{f}_{s}}{2r^{4}f_{s}^{2}}}{\zeta_{s}^{2}-1},\label{eq:phis}
\end{align}
where $f_{s}=f(\phi_{s})$, $\dot{f}_{s}=\frac{df(\phi_{s})}{d\phi_{s}}$
and $\alpha_{s},\zeta_{s},\phi_{s}$ are functions independent of
time. Their asymptotic behaviors at spatial infinity are given by
$\phi_{s}\to\frac{Q_{s}}{r},$ $\zeta_{s}\to\sqrt{\frac{2M}{r}},$
and $\alpha_{s}\to1.$ Here $M$ is the total mass and $Q_{s}$ the
scalar charge of the static black hole solution. Near the horizon
$r=r_{h}$, the fields behave as $\phi_{s}\to\phi_{h},\zeta_{s}\to1,\alpha_{s}\to\alpha_{h},$
with certain  values $\phi_{h},\alpha_{h}$. For fixed electric charge $Q$
and horizon radius $r_{h}$, the static solution can be obtained by
adjusting $\phi_{h}$ using shooting method. The total mass and
scalar charge are determined from the asymptotic behavior of the fields:
$M=\lim_{r\to\infty}\frac{r}{2}\zeta_{s}^{2}$ and $Q_{s}=-\lim_{r\to\infty}r^{2}\partial_{r}\phi_{s}$.
These parameters are not independent but related. For example, they
satisfy the Smarr relation $M^{2}+Q_{s}^{2}=Q^{2}+\frac{r_{h}^{2}\alpha_{h}^{2}}{4}\left(1-\frac{Q^{2}}{f(\phi_{h})r_{h}^{2}}\right).$
For $f=e^{100\phi^{4}}$, the static solution space is shown in Fig.
\ref{fig:static} in the main text. The bifurcation
structure is similar for other values of the coupling parameter $\lambda$.

\textbf{Dynamical evolution}. 
We perturb a stable hairy black hole $\alpha_{s},\zeta_{s},\phi_{s}$
with an ingoing scalar pulse $\delta\phi(r)$. The initial configuration
of the scalar field is
\begin{align}
\phi_{in} & =\phi_{s}(r)+\delta\phi(r),\\
\Pi_{in} & =\Pi_{s}(r)+\frac{1}{\alpha_{s}}\partial_{r}\delta\phi(r),
\end{align}
where $\Pi_{s}=-\zeta_{s}\partial_{r}\phi_{s}$. This choice can minimize
the outgoing wave and thus improve numerical accuracy for a long time
numerical simulation. Given $\phi_{in}$ and $\Pi_{in}$, the initial
values of the metric functions $\zeta$ and $\alpha$ can be calculated
 by (\ref{eq:zetadr}) and (\ref{eq:alphadr}), respectively. Using
(\ref{eq:zetadt}-\ref{eq:Pt}), we get the $\zeta,\phi,\Pi$ on the
next time slice, and $\alpha$ can be worked out through (\ref{eq:alphadr}).
Repeating this procedure, we can obtain the metric and scalar functions
at all time steps. 

We compact the radial coordinate by $z=\frac{r}{r+k}$ where $k$
is constant. The system is evolved in the $z$ coordinate, ranging
from $z_{c}=\frac{r_{c}}{r_{c}+k}$ to $z=1$. The position $r=r_{c}$
locates close to the initial apparent horizon from the interior. Since
PG coordinates are horizon penetrate, this makes the numerics works
well. We typically
discretize $z$ uniformly using $2^{11}$ grid points and
employ the fourth order finite difference method in the radial direction
and the fourth-order Runge-Kutta method in the time direction. To
stabilize the simulation, we adopt the Kreiss-Oliger dissipation.
In the first step, the Newton-Raphson method is used to solve the
constraint equation (\ref{eq:zetadr}). The convergence is estimated
by $\frac{u_{2N}-u_{N}}{u_{4N}-u_{2N}}=2^{n}+O(1/N)$ where $u_{N}$
is the results obtained by $N$ grid points and $n$ is the accuracy
order. It turns out that our numerical solutions converge to the fourth
order.

In PG coordinates, the apparent horizon during evolution is located
where $\zeta(t,r_{h})=1$. Then the scalar field value on the
horizon can be simply obtained by $\phi_{h}(t)=\phi(t,r_{h})$. Typical
evolution results are shown in Fig. \ref{fig:bottleEvo} in the main
text. All near-critical simulations obey the power-law  $t_{b}\propto\epsilon^{-1/4}$
for the bottleneck duration.

For spherical, non-static spacetimes, the quasi-local Misner-Sharp
mass is often used to measure the total energy within a spherical
surface \citep{Hayward:1994bu}. It is defined as
\begin{equation}
m(t,r)=\frac{r}{2}\left(1-g^{\mu\nu}\nabla_{\mu}r\nabla_{\nu}r\right),
\end{equation}
where $r$ is the areal radius. In asymptotically flat spacetimes
it reduces to the ADM mass at spatial infinity and to the Bondi-Sachs
mass at future null infinity. In PG coordinates, it is easy to derive
 $m(t,r)=\frac{r}{2}\zeta^{2}$ and the outgoing radial energy flux,
\begin{equation}
-\frac{dm(t,r)}{dt}=-r^{2}\alpha\left(\Pi+\zeta\partial_{r}\phi\right)\left(\Pi\zeta+\partial_{r}\phi\right),
\end{equation}
since the energy reduction per unit of time is the outgoing energy flux.

\subsection*{The asymptotical analysis for bottleneck}

To understand the origin of the power-law  observed in the
numerical simulations, we perform a systematic asymptotic analysis
around the bifurcation solution. This approach, rooted in the principles
of center manifold reduction \citep{Keener2000,kuznetsov1998elements}, allows us to isolate the slow dynamics
associated with the zero mode.  This derivation is largely independent of the model details, but depends on the bifurcation structure of the solution space.  By deriving an effective amplitude
equation for this mode, we provide an analytical explanation for the
universal critical behavior near the bifurcation point.

For our analysis, we adopt the same PG coordinates used in the numerical
simulations. Rather than introducing auxiliary variables, we work
directly with the metric functions $\zeta(t,r)$, $\alpha(t,r)$ and
the scalar field $\phi(t,r)$, which satisfy the full nonlinear field
equations:
\begin{align}
\partial_{t}\zeta= & r\partial_{t}\phi\left(\frac{\partial_{t}\phi}{\alpha}-\frac{(\zeta{}^{2}-1)\partial_{r}\phi}{\zeta}\right),\\
\partial_{r}\alpha= & r\partial_{r}\phi\left(\alpha\partial_{r}\phi-\frac{\partial_{t}\phi}{\zeta}\right),\\
\partial_{r}\zeta= & \frac{r}{2\zeta}\left(\frac{(\partial_{t}\phi)^{2}}{\alpha{}^{2}}+(\partial_{r}\phi)^{2}\right)+\frac{Q^{2}}{2r^{3}\zeta f(\phi)}\\
 & -\frac{\zeta}{2}\left(r(\partial_{r}\phi)^{2}+\frac{1}{r}\right),\nonumber \\
\partial_{t}^{2}\phi= & \left(\frac{r(\partial_{t}\phi)^{2}}{2\alpha\zeta}+\frac{Q^{2}\alpha}{2r^{3}\zeta f(\phi)}\partial_{t}\phi\right)\partial_{t}\phi\\
 & +\left(\frac{\partial_{t}\alpha}{\alpha}+\frac{3\alpha\zeta}{2r}-\frac{r\alpha\left(\zeta^{2}-1\right)(\partial_{r}\phi)^{2}}{2\zeta}\right)\partial_{t}\phi\nonumber \\
 & +2\alpha\zeta\partial_{t}\partial_{r}\phi-\alpha^{2}\left(\zeta^{2}-1\right)\partial_{r}^{2}\phi\nonumber \\
 & -\alpha^{2}\frac{\partial_{r}\phi}{r}\left(\zeta^{2}-2+\frac{Q^{2}}{r^{2}f(\phi)}\right)+\frac{Q^{2}\alpha^{2}f'(\phi)}{2r^{4}f(\phi)^{2}}.\nonumber 
\end{align}
Near the bifurcation point, the system exhibits a saddle-node structure
in solution space, where two branches of static hairy black hole solutions
merge. Linear perturbation theory reveals that the bifurcation solution
is characterized by a dominant zero mode, an eigenmode of the
linearized equations with a vanishing frequency. Perturbations along
this mode experience critical slowing down, evolving on time scales
much longer than those associated with other modes.  

To systematically resolve this slow dynamics, we introduce a rescaled
time variable $T=\varepsilon^{1/4}t$ where $\varepsilon=M/M_{*}-1$
is a small parameter quantifying the deviation from the bifurcation
point. The mass deviation $\varepsilon$ is different to the perturbation
deviation $\epsilon$ introduced earlier, but it is approximately
proportional to $\epsilon$. The choice of the exponent $1/4$ is
crucial: it ensures that the inertial term $\partial_{t}^{2}\phi$
contributes to the final amplitude equation at the same order as the
leading-order nonlinear terms, thus guaranteeing a non-trivial dynamics.
This rescaling magnifies the evolution of the soft mode, which would
otherwise appear frozen on the original time scale.

With this preparation, we expand the fields $\varPsi=\{\phi,\zeta,\alpha\}$
in the following series:
\begin{equation}
\varPsi\to\varPsi_{*}(r)+\varepsilon^{\frac{1}{2}}A\varPsi_{0}(r)+\varepsilon^{\frac{3}{4}}\frac{dA}{dT}\varPsi_{l}(r)+\varepsilon\varPsi_{n}(T,r).\label{eq:Psi0ln}
\end{equation}
The first term $\varPsi_{*}$ is the static background solution at
the bifurcation point, serving as the reference configuration from
which the perturbation departs. The second term is the leading-order
perturbation along the zero mode $\varPsi_{0}$, with slowly varying
amplitude $A(T)$ that encodes the effective one-dimensional dynamics
near the bifurcation point. The characteristic $\varepsilon^{1/2}$
scaling is universal near a saddle-node bifurcation, reflecting the
fact that the deviation of the solution from the bifurcation point
grows as the square root of the control parameter. Mathematically,
it ensures that the leading nonlinear term, proportional to $A^{2}$,
enters at the same order as the other dominant terms in the final
amplitude equation. The third term $\varPsi_{l}$ is not a new independent
mode but represents the linear response to the slow drift of the zero
mode amplitude. Its inclusion is a necessary ingredient for the asymptotic
expansion to be self-consistent. The fourth term $\varPsi_{n}$ describes
the nonlinear response to the zero mode. Importantly, $\varPsi_{n}$
is a nonlinear functional of $A$ and its derivatives. It is at this
order that the key physical effects, nonlinearity and inertia, compete
directly, ultimately resulting the bottleneck phenomenon. 

It is instructive to contrast this procedure with the standard perturbative
analysis of QNMs. In that approach, one expands around a static background
and solves for modes with harmonic time dependence $e^{-i\omega t}.$
For the zero mode, $\omega=0,$ meaning its amplitude is time-independent.
In our framework, we promote the zero mode amplitude to a dynamical
variable $A$ that evolves on a slow time $T$. When $A$ is taken
to be constant, our expansion reduces to the standard static perturbation
theory and agrees with the QNM result. However, when $A$ varies slowly,
the dynamics becomes intrinsically nonlinear in time and captures
the bottleneck behavior, a phenomenon that is inaccessible to the
QNM analysis. We emphasize that the amplitude $A$ is not free
but must satisfy certain dynamic constraint to keep the expansion (\ref{eq:Psi0ln})
self-consistency. In fact, it is this constraint equation governing
$A$ that unveils deeper physical insights. 

Substituting the asymptotic expansion (\ref{eq:Psi0ln}) into the
full field equations and collecting terms order by order, we get a
hierarchy of equations. At order $\varepsilon^{0}$, we get the equations
for the static background solution. At order $\varepsilon^{1/2}$,
the equations govern the zero mode; at order $\varepsilon^{3/4}$,
the linear response mode; and at order $\varepsilon$, the nonlinear
response, which ultimately yields an ordinary differential equation
for the amplitude $A$. The background fields $\phi_{*},\zeta_{*},\alpha_{*}$
at the bifurcation point satisfy the $\varepsilon^{0}$-order equations,
which have the form in (\ref{eq:alphas}-\ref{eq:phis}). The asymptotic
behavior at spatial infinity is given by $\phi_{*}\to\frac{Q_{*}}{r},$
$\zeta_{*}\to\sqrt{\frac{2M_{*}}{r}}$ where $M_{*}$ is the total
mass and $Q_{*}$ the scalar charge of the hairy solution at the 
bifurcation point. 

By introducing the coordinate transformation $dr_{*}=\frac{1}{\left(1-\zeta_{*}^{2}\right)\alpha_{*}}dr$,
and defining the variables $\psi_{0}=r\phi_{0}$, $\psi_{l}=r\phi_{l}$,
$\psi_{n}=r\phi_{n}$, we find that the zero mode $\psi_{0}$, the
linear response $\psi_{l}$ and nonlinear response $\psi_{n}$ satisfy
the following equations derived from the $\varepsilon^{1/2}$, $\varepsilon^{3/4}$,
and $\varepsilon$ orders of the expansion, respectively: 
\begin{align}
0= & \partial_{*}^{2}\psi_{0}+U\psi_{0},\label{eq:psi0rr}\\
0= & \partial_{*}^{2}\psi_{l}+U\psi_{l}+S_{l},\label{eq:psilrr}\\
0= & \partial_{*}^{2}\psi_{n}+U\psi_{n}+S_{n}.\label{eq:psinrr}
\end{align}
Here $\partial_{*}=\frac{\partial}{\partial r_{*}}$, and the effective
potential $U$ is determined by the background solution,
\begin{align}
\frac{U}{\left(\zeta_{*}^{2}-1\right)\alpha_{*}^{2}}= & \frac{\zeta_{*}^{2}}{r^{2}}+2\left(\frac{Q^{2}}{r^{2}f_{*}}-1\right)\left(\partial_{r}\phi_{*}\right)^{2}\\
 & +\frac{Q^{2}}{r^{2}f_{*}}\left(\frac{2\dot{f}_{*}^{2}-\ddot{f}_{*}f_{*}}{2r^{2}f_{*}^{2}}-\frac{2\partial_{r}f_{*}}{f_{*}r}-\frac{1}{r^{2}}\right),\nonumber 
\end{align}
with $\ddot{f}_{*}=\frac{d^{2}f(\phi_{*})}{d\phi_{*}^{2}}$. The source
term $S_{l}$ for $\psi_{l}$ is a linear function of the zero mode,
\begin{align}
\frac{S_{l}}{\alpha_{*}\left(1-\zeta_{*}^{2}\right)}= & \left(r^{2}\left(1-\zeta_{*}^{2}\right)\left(\partial_{r}\phi_{*}\right)^{2}+3\zeta_{*}^{2}\right)\frac{\phi_{0}}{2\zeta_{*}}\\
 & +\frac{Q^{2}}{r^{2}f_{*}}\frac{\phi_{0}}{2\zeta_{*}}+2r\zeta_{*}\partial_{r}\phi_{0}.\nonumber 
\end{align}
It is time-independent. The source term $S_{n}=X\frac{d^{2}A}{dT^{2}}+YA^{2}$
for $\psi_{n}$ is time-dependent through the amplitude $A$, in which
the coefficients $X$ and $Y$ are given by
\begin{align}
\frac{X}{\alpha_{*}\left(1-\zeta_{*}^{2}\right)}= & \left(r^{2}\left(1-\zeta_{*}^{2}\right)\left(\partial_{r}\phi_{*}\right)^{2}+3\zeta_{*}^{2}\right)\frac{\phi_{l}}{2\zeta_{*}}\\
 & +\frac{\phi_{l}}{2\zeta_{*}}\frac{Q^{2}}{r^{2}f_{*}}+2r\zeta_{*}\partial_{r}\phi_{l}-\frac{r\phi_{0}}{\alpha_{*}},\nonumber 
\end{align}
\begin{align}
\frac{Y}{\alpha_{*}^{2}\left(1-\zeta_{*}^{2}\right)}= & 2r^{2}\left(1-\frac{Q^{2}}{r^{2}f_{*}}\right)\phi_{0}^{2}\left(\partial_{r}\phi_{*}\right){}^{3}\\
 & +\frac{3Q^{2}\phi_{0}^{2}}{2f_{*}}\left(\frac{2\partial_{r}f_{*}}{rf_{*}}+\frac{f_{*}\ddot{f}_{*}-2\dot{f}_{*}^{2}}{r^{2}f_{*}^{2}}\right)\partial_{r}\phi_{*}\nonumber \\
 & +\frac{Q^{2}\phi_{0}^{2}}{4r^{3}f_{*}^{4}}\left(f_{*}^{2}\dddot{f}_{*}+6\dot{f}_{*}^{3}-6f_{*}\dot{f}_{*}\ddot{f}_{*}\right)\nonumber \\
 & +\frac{3}{2}\partial_{r}\phi_{0}^{2}\left(\frac{Q^{2}\dot{f}_{*}}{2r^{2}f_{*}^{2}}+r\left(1-\frac{Q^{2}}{r^{2}f_{*}}\right)\partial_{r}\phi_{*}\right).\nonumber 
\end{align}
Here $Y$ depends only on the background and the zero mode, while
$X$ also depends on the linear response. 

According to the Fredholm alternative theorem \citep{Keener2000}, a solution to the inhomogeneous
equation (\ref{eq:psinrr}) exists if and only if the source term
is orthogonal to the kernel of the adjoint operator of $L=\partial_{*}^{2}+U$.
In the present case, the linear operator $L$ acting on $\psi_{n}$
in (\ref{eq:psinrr}) is identical to that of the homogeneous equation
(\ref{eq:psi0rr}) governing the zero mode $\psi_{0}$. Since $L$
is self-adjoint with unit weight function $\omega=1$, its kernel is spanned
precisely by the zero mode $\psi_{0}$. Therefore, the solvability
condition for the equation of $\psi_{n}$ is obtained by multiplying
(\ref{eq:psinrr}) by $\psi_{0}$ and integrating over the entire
exterior region:
\begin{equation}
\int_{-\infty}^{\infty}\psi_{0}S_{n}dr_{*}=-\int_{-\infty}^{\infty}\psi_{0}\left(\partial_{*}^{2}\psi_{n}+U\psi_{n}\right)dr_{*}.
\end{equation}

The right-hand side can be simplified via integration by parts. Using
equation (\ref{eq:psi0rr}) for $\psi_{0}$, it reduces to a boundary
term involving the Wronskian $w=\left.\left(1-\zeta_{*}^{2}\right)\alpha_{*}r^{2}\left(\phi_{0}\partial_{r}\phi_{n}-\phi_{n}\partial_{r}\phi_{0}\right)\right|_{r_{h}}^{\infty}.$
Consequently, the solvability condition takes the form of a one-dimensional
ordinary differential equation for the amplitude 
\begin{equation}
0=a\frac{d^{2}A}{dT^{2}}+bA^{2}+w,\label{eq:ATT0}
\end{equation}
where the coefficients are given by the bulk integrals $a=\int_{-\infty}^{\infty}\psi_{0}Xdr_{*},$
$b=\int_{-\infty}^{\infty}\psi_{0}Ydr_{*}$. This condition ultimately
governs the dynamics of the amplitude $A$. Above procedure effectively
constitutes a center manifold reduction: we project the full infinite-dimensional
field equations onto the one-dimensional subspace spanned by the zero
mode, ultimately yielding an effective amplitude equation for a nonlinear
oscillator. 

However, the situation is complicated by the fact that both the boundary
term $w$ and the integral $a$ diverge. This divergence arises due
to the slow fall-off of the fields at spatial infinity and must be
carefully regularized to extract finite and physically meaningful
dynamics.

To evaluate the boundary Wronskian, it is necessary to determine the
asymptotic behaviors of $\phi_{0}$, $\phi_{l}$ and $\phi_{n}$ near
both the horizon and spatial infinity. All functions remain regular
at the horizon thanks to the choice of PG coordinates, which eliminate
the coordinate singularity there. But the behavior near spatial infinity
is more subtle and requires careful analysis. Using series expansion
at spatial infinity, we find the zero mode behaves as $\phi_{0}\to\frac{q_{0}}{r}$,
where $q_{0}$ is a normalization constant, conveniently set to $q_{0}=1$.
This zero mode induces a linear response in $\phi_{l}$, which asymptotically
behaves as $\phi_{l}\to-2q_{0}\sqrt{\frac{2M_{*}}{r}}$. The nonlinear
response $\phi_{n}$ requires a more careful treatment. Since $\phi_{n}$
satisfies a linear inhomogeneous equation (\ref{eq:psinrr}), it is
convenient to decompose it into three distinct contributions, 
\begin{equation}
\phi_{n}(r,T)=\rho(r)\frac{d^{2}A}{dT^{2}}+\chi(r)A^{2}+\gamma(r),
\end{equation}
where the time-independent functions $\rho$, $\chi$, $\gamma$ encode
different physical responses. Specifically, $\rho$ encodes the inertial
response to the acceleration of the zero-mode amplitude, $\chi$ encodes
the nonlinear response to the self-interaction of the zero mode, and
$\gamma$ encodes the static response independent of $A$ which corresponds
to a static shift of the background black hole parameters (e.g. $M\to M_{*}(1+\varepsilon)$). 

Substituting this decomposition into the inhomogeneous equation (\ref{eq:psinrr})
leads to three decoupled equations for $\rho,$ $\chi$ and $\gamma$.
Among these, $\chi\to-\frac{q_{0}^{2}Q_{*}}{2r^{3}}$ decays rapidly
at infinity and is irrelevant for the solvability condition. The dominant
asymptotic behaviors are therefore
\begin{equation}
\rho\to q_{0}\left(\frac{1}{2}r+4M_{*}\log\frac{r}{r_{h}}\right),\ \ \ \gamma\to q_{m},\label{eq:qm}
\end{equation}
where $q_m$ is a parameter to be determined.

With these asymptotics, the boundary Wronskian receives its entire
contribution from spatial infinity:
\begin{equation}
w=q_{0}^{2}\left[\left(r+4M_{*}\ln\frac{r}{r_{h}}\right)_{r\to\infty}+\frac{7}{2}M_{*}\right]\frac{d^{2}A}{dT^{2}}+q_{m}q_{0}.\label{eq:w}
\end{equation}
The divergence here stems from the inertial response $\rho A''(T)$.
A similar divergence also appears in the bulk integral $a$, whose
asymptotic behavior is $a\to-q_{0}^{2}\int^{\infty}\left(1+\frac{4M_{*}}{r}+\cdots\right)dr.$
Crucially, these divergences do not invalidate the analysis. The key
insight is that the divergent term in the boundary Wronskian $w$
is exactly canceled by a corresponding divergence in the bulk integral
$a$. To make this cancellation explicit, we define a regularized
coefficient 
\begin{align}
a_{0} & =a+q_{0}^{2}\int_{r_{h}}^{\infty}\left(1+\frac{4M}{r}\right)dr\label{eq:a0}\\
 & =a+q_{0}^{2}\left[\left(r+4M\ln\frac{r}{r_{h}}\right)_{r\to\infty}-r_{h}\right].\nonumber 
\end{align}
Substituting (\ref{eq:w}) and (\ref{eq:a0}) into the solvability
condition (\ref{eq:ATT0}), the divergent terms cancel exactly, leaving
a regular amplitude equation,
\begin{equation}
0=\left[\frac{a_{0}}{q_{0}}+q_{0}\left(\frac{7}{2}M_{*}+r_{h}\right)\right]\frac{d^{2}A}{dT^{2}}+\frac{b}{q_{0}}A^{2}+q_{m}.\label{eq:ATT1}
\end{equation}
This simple equation governs the slow dynamics near the bifurcation
point. 

The successful regularization clarifies the meaning of the divergence
in $\rho$ as $r\to\infty$. It is not a physical singularity, but
a coordinate artifact stemming from the asymptotic time foliation
in PG coordinates. While a geometric regularization, such as transitioning
to hyperboloidal slicing, would avoid such divergences from the outset
\citep{PanossoMacedo:2024nkw}, the present analysis demonstrates
that even within PG coordinates, the Fredholm alternative guarantees
that all divergent contributions cancel exactly in the solvability
condition. This algebraic regularization robustly isolates the physical
dynamics, yielding a well-posed amplitude equation. The term $\rho A''(T)$
should thus be understood as describing the transient inertial response
of the zero mode, which influences the time-dependent dynamics but
does not alter the background parameters like the black hole mass.
In contrast, the static part $\gamma$ induces a well-defined mass
modification \citep{Konoplya:2011qq}. Based on the asymptotic expansion (\ref{eq:Psi0ln})
for the metric function $\zeta$, we find that the lower-order metric
perturbations $\zeta_{0}\to-q_{0}\sqrt{\frac{2M_{*}}{r}}\frac{Q_{s}}{2M_{*}r}$
and $\zeta_{l}\to q_{0}\frac{2Q_{s}}{r}$, while the static response
$\zeta_{n}\to-\sqrt{\frac{2M_{*}}{r}}\frac{Q_{*}q_{m}}{2M_{*}}$ sourced
by $\gamma$ dominates. This behavior implies a static mass shift
given by $M=\lim_{r\to\infty}\frac{r}{2}\zeta^{2}\to M_{*}-\varepsilon q_{m}Q_{*}$.
Meanwhile, since our entire asymptotic analysis is built on the assumption
that the mass changes as $M\to M_{*}(1+\varepsilon)$, consistency
requires $q_{m}=-\frac{M_{*}}{Q_{*}}.$ Thus the static response encodes
a physical transition between two black hole solutions with different
masses. 

With all coefficients determined, we can analyze the final equation
of motion. We emphasize that the dynamics described by (\ref{eq:ATT1})
are independent of the normalization $q_{0}$. As seen in (\ref{eq:Psi0ln}),
the physical perturbation appears in the combination $\sqrt{\varepsilon}A(T)\phi_{0}(r)$.
Rescaling $q_{0}\to Kq_{0}$ with $A\to A/K$ leaves the dynamics
invariant, since $a_{0}\to K^{2}a_{0}$, $b\to K^{3}b$, and the ODE
(\ref{eq:ATT1}) remains unchanged. For numerical convenience we simply
set $q_{0}=1$. Then the numerics show that the zero mode $\phi_{0}$
is positive in the whole space. Introducing the variable $\Lambda=\varepsilon^{1/2}A$
and recalling that $t\to\varepsilon^{-1/4}T$, the amplitude equation
takes the form of an underdamped nonlinear oscillator,
\begin{equation}
\frac{d^{2}\varLambda}{dt^{2}}=-\mu\varepsilon-\beta\varLambda^{2},\label{eq:Lt}
\end{equation}
where $\mu=-\frac{M_{*}}{Q_{*}\left(a_{0}+\frac{7}{2}M_{*}+r_{h}\right)}$
and $\beta=-\frac{b}{a_{0}+\frac{7}{2}M_{*}+r_{h}}$, both determined
by the property of the static solution at the bifurcation point. The
numerics show that $\mu$ and $\beta$ are positive. For negative
$\varepsilon$, the equation admits two equilibria $\Lambda=\pm\sqrt{-\mu\varepsilon/\beta}$,
one stable and another unstable, which meet and annihilate as $\varepsilon$
becomes positive. This is the hallmark of a saddle-node bifurcation.
For small positive $\varepsilon$, the system can evolve very slow
as if it were attracted by the saddle-node ghost 
before accelerating away. So above equation captures the essence of
the bottleneck effect.

Finally, we emphasize the scope of validity of this derivation. The
nonlinear oscillator captures the dynamics asymptotically close to
the bifurcation, where the zero mode dominates and higher-order corrections
remain subleading. Far from the bifurcation, or once the amplitude
grows large enough that neglected terms compete with leading nonlinearities,
this effective model ceases to apply and the full field dynamics must
be considered. 

%\end{linenumbers}

\end{document}